\documentclass[twocolumn,showpacs,amsmath,nobibnotes,nofootinbib,floatfix,nonbibnotes,nofootinbib,notitlepage,longbibliography]{revtex4-1}
\usepackage{amsmath}
\usepackage{amssymb}
\usepackage{bm}
\usepackage{epsfig}
\usepackage{graphicx}
\usepackage[english]{babel}
\usepackage[unicode=true,colorlinks=true,citecolor=blue,urlcolor=blue]{hyperref}

\begin{document}

\title{Intrinsic exciton states mixing and non-linear optical properties in transition metal dichalcogenide monolayers}

\author{M.\,M.\,Glazov}\thanks{Corresponding author: glazov@coherent.ioffe.ru}
\author{L.\,E.\,Golub}
\affiliation{Ioffe Institute, St.~Petersburg 194021, Russia}
\author{G. Wang}
\author{X. Marie}
\author{T. Amand}
\author{B. Urbaszek}
\affiliation{Universit\'{e} de Toulouse, INSA-CNRS-UPS, LPCNO, 31077 Toulouse, France}

\date{\today}

\begin{abstract}
Optical properties of transition metal dichalcogenides monolayers are controlled by the Wannier-Mott excitons forming a series of $1s$, $2s$, $2p$, \ldots hydrogen-like states. We develop the theory of the excited excitonic states energy spectrum fine structure. We predict that $p$- and $s$-shell excitons are mixed due to the specific $D_{3h}$ point symmetry of the transition metal dichalcogenide monolayers. Hence, both $s$- and $p$-shell excitons are active in both single- and two-photon processes providing an efficient mechanism of second harmonic generation. The corresponding contribution to the nonlinear susceptibility is calculated.
\end{abstract}


\maketitle

\section{Introduction}

Transition metal dichalcogenides monolayers (TMD MLs) such as MoS$_2$, MoSe$_2$, WS$_2$, WSe$_2$, etc., are prototypical two-dimensional (2D) semiconductors with direct band gaps of the order of 2~eV at the Brillouin zone edges~\cite{Mak:2010bh,Xiao:2012cr,Wang:2012aa,C4CS00301B,Kolobov2016book}.
Optical properties of these systems are mainly governed by the Coulomb correlated electron-hole pairs, excitons~\cite{Yu30122014,Moody:16}. Due to relatively large effective masses of the electrons and holes and rather weak screening of the Coulomb interaction in 2D systems~\cite{chaplik_entin,1979JETPL..29..658K} excitons are highly stable in TMD MLs: The predicted~\cite{Cheiwchanchamnangij:2012pi,Ramasubramaniam:2012qa,PhysRevB.87.155304,Qiu:2013fe,PhysRevB.89.125309} and observed~\cite{Mak:2013lh,PhysRevLett.113.026803,Chernikov:2014a,Ye:2014aa,Klots:2014aa,Wang:2015b} exciton binding energies amount to $\sim 500$~meV. The oscillator strengths of excitons in TMD MLs are also much higher as compared with conventional 2D structures based, e.g., on GaAs~\cite{KornMoS2,PhysRevB.90.075413,doi:10.1021/nl503799t,Poellmann:2015aa}. Just like the direct Coulomb interaction, the electron-hole exchange interaction in TMD MLs is also stronger than in conventional semiconductors~\cite{glazov2014exciton,PhysRevB.89.205303,Yu:2014fk-1,PhysRevLett.115.176801}. It controls the spin and valley dynamics of excitons in a wide range of temperatures~\cite{PhysRevB.90.161302}, see for review Ref.~\cite{PSSB:PSSB201552211} and references therein.

Excitons also play a crucial role in nonlinear optical properties, particularly, in two-photon absorption and second harmonic generation in 2D TMD~\cite{PhysRevLett.113.026803,Ye:2014aa,Wang:2015b,Seyler:2015aa,PhysRevB.92.161409,PhysRevB.89.235410}. {The nonlinear susceptibility is enhanced in the vicinity of excitonic resonances due to redistribution of the oscillator strength from a continuum to bound electron-hole pair states~\cite{Wang:2015b,PhysRevB.89.235410,PhysRevB.89.081102}.} It is usually assumed that the $s$-shell exciton states are active in single-photon processes, while $p$-shell excitons are optically active in two-photon processes~\cite{ivchenko05a,PhysRevB.92.085413}. The energies of the ground, $1s$, and excited $2s$, $2p$, \ldots excitonic states strongly differ from the hydrogenic series due to unusual screening in TMD MLs~\cite{Chernikov:2014a}.  Moreover, for the same reason, in TMD MLs the ``accidental'' Coulomb degeneracy of excited excitonic states is lifted. As a result, the fine structure of the excited, e.g., $2p$-exciton states in TMD MLs has become topical nowadays, in particular, in view of possible manifestations of the Berry curvature effect~\cite{2p:1,2p:2}.

Here we demonstrate that the $s$- and $p$-shell excitonic states are mixed in TMD monolayers. We present the detailed theory of the fine structure of excited excitonic states in TMD MLs, including the symmetry analysis and the microscopic theory.  The microscopic mechanisms of the mixing are uncovered and the detailed model is developed within the $\bm k\cdot \bm p$ formalism. The mixing is shown to be quite substantial, for example, the coupling constant of $2s$- and $2p$-shell states is estimated to be in the range from tenths to several meV. In particular, the mixing makes $s$-shell states active in the two-photon absorption and also contributes to the second harmonic generation in TMD MLs. The excitonic contributions to the two-photon absorption and nonlinear susceptibility responsible for the second harmonic generation are calculated.


\section{Symmetry analysis}\label{sec:symmetry}

The exciton basic wavefunctions in semiconductors can be recast as products of the two-particle envelope and Bloch functions. For the exciton freely propagating in the TMD ML plane with the center of mass wavevector $\bm K$ the wavefunction $\Psi$ reads~\cite{PSSB:PSSB201552211}
\begin{equation}
\label{ex:gen}
\Psi_{\bm K;\nu,\mu} (\bm \rho_e, \bm \rho_h) = {\exp{(\mathrm i \bm K \bm R)}} \Phi_\nu(\bm \rho) \mathcal U_{\mu}(\bm \rho_e,\bm \rho_h).
\end{equation}
The exciton state is characterised by quantum numbers $\nu=1s, 2s, 2p, \ldots$ denoting the hydrogen-like relative motion states and $\mu$ enumerating the band states, respectively, $\Phi_\nu(\bm \rho)$ is the relative motion wavefunction and $\mathcal U_{\mu}(\bm \rho_e,\bm \rho_h)$ is the two-particle Bloch function. Here $\bm \rho_e$, $\bm \rho_h$ are the electron and hole in-plane coordinates, respectively, $\bm \rho = \bm \rho_e - \bm \rho_h$ is the relative coordinate, $\bm R=(m_e \bm \rho_e + m_h \bm \rho_h)/(m_e+m_h)$ is the center of mass position vector, $m_e$ ($m_h$) are the electron (hole) effective masses, the normalization area is set to unity. In what follows we focus on the states with  $\bm K=0$ and restrict ourselves with optically active Bloch states denoted as $\mu=\pm 1$ (active in $\sigma^\pm$ polarizations{, respectively}) or, alternatively, as $\mu=x,y$, where $x$ and $y$ denote orientation of oscillating dipole moment in the Cartesian coordinate frame with the in-plane axes $x$, $y$ and the normal to the TMD ML~$z$. The ``dark'' excitons with opposite spins of the electron and the hole which are weakly optically active in the polarization perpendicular to the ML plane are disregarded~\cite{glazov2014exciton,PhysRevB.93.121107,2016arXiv160302572S}.

\begin{figure}[h]
\includegraphics[width=0.6\linewidth]{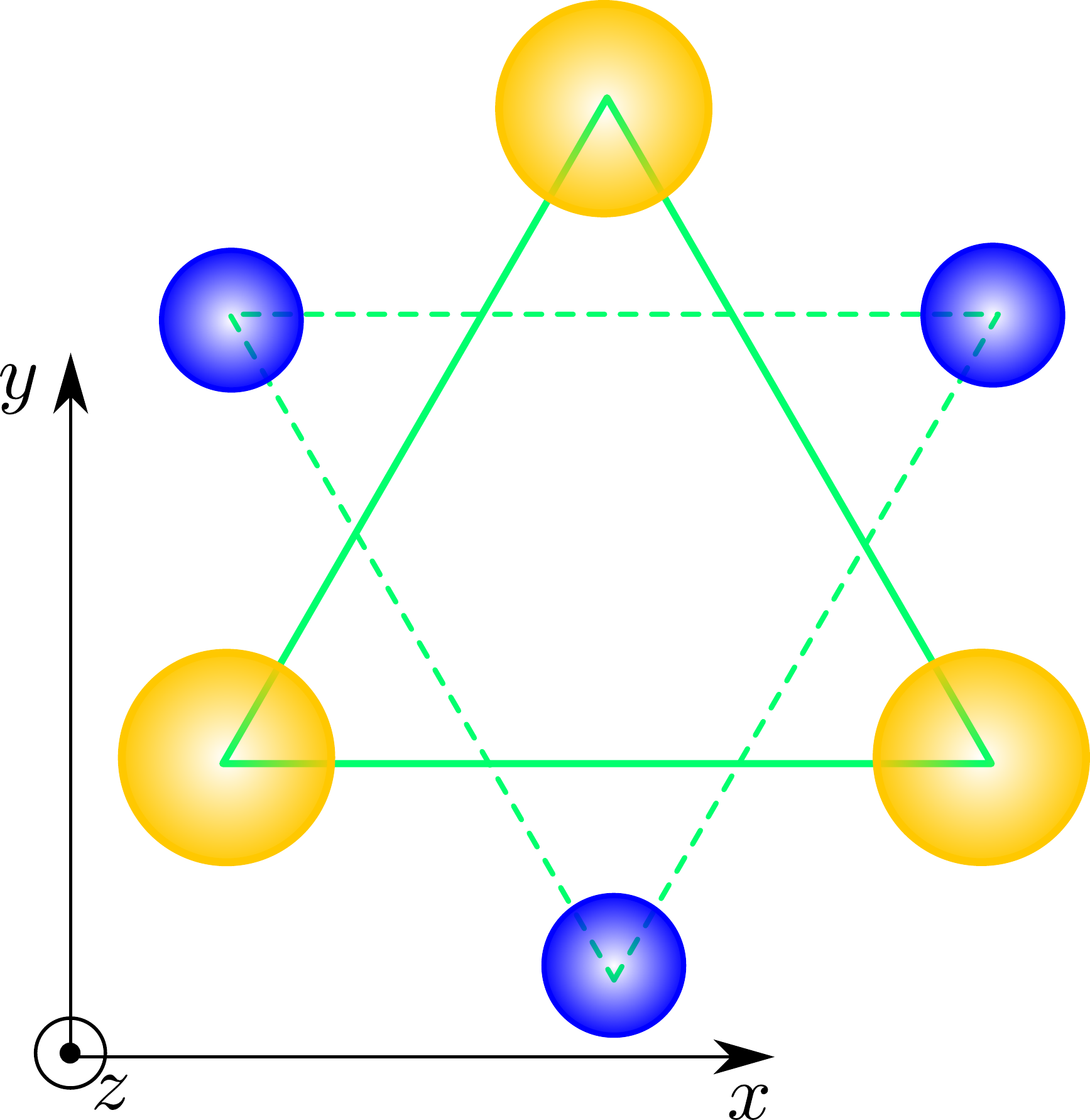}
\caption{Top view of atomic arrangement in TMD ML. Yellow circles show positions of the transition metal atoms, blue circles show the projection of chalcogen atoms on the horizontal plane. Green lines are guides for eye showing non-equivalent triangles responsible for the overall $D_{3h}$ point symmetry of the ML.}\label{fig:atoms}
\end{figure}

The point symmetry of the TMD ML is $D_{3h}$, in the chosen coordinate frame $z$ is the three-fold rotation axis and $y$ is (one of three) two-fold in-plane rotation axes, the horizontal reflection plane $\sigma_h\parallel (xy)$, one of three vertical reflection planes is $\sigma_v\parallel (yz)$ and two remaining planes can be obtained from $\sigma_v$ by the rotation through the angles $\pm 2\pi/3$ around $z$-axis. The set of axis and atomic arrangement is illustrated in Fig.~\ref{fig:atoms}. In accordance with the general theory, the exciton eigenstates in TMD ML transform according to irreducible representations of the $D_{3h}$ point group~\cite{ivchenko05a}.

\begin{figure}[h]
\includegraphics[width=\linewidth]{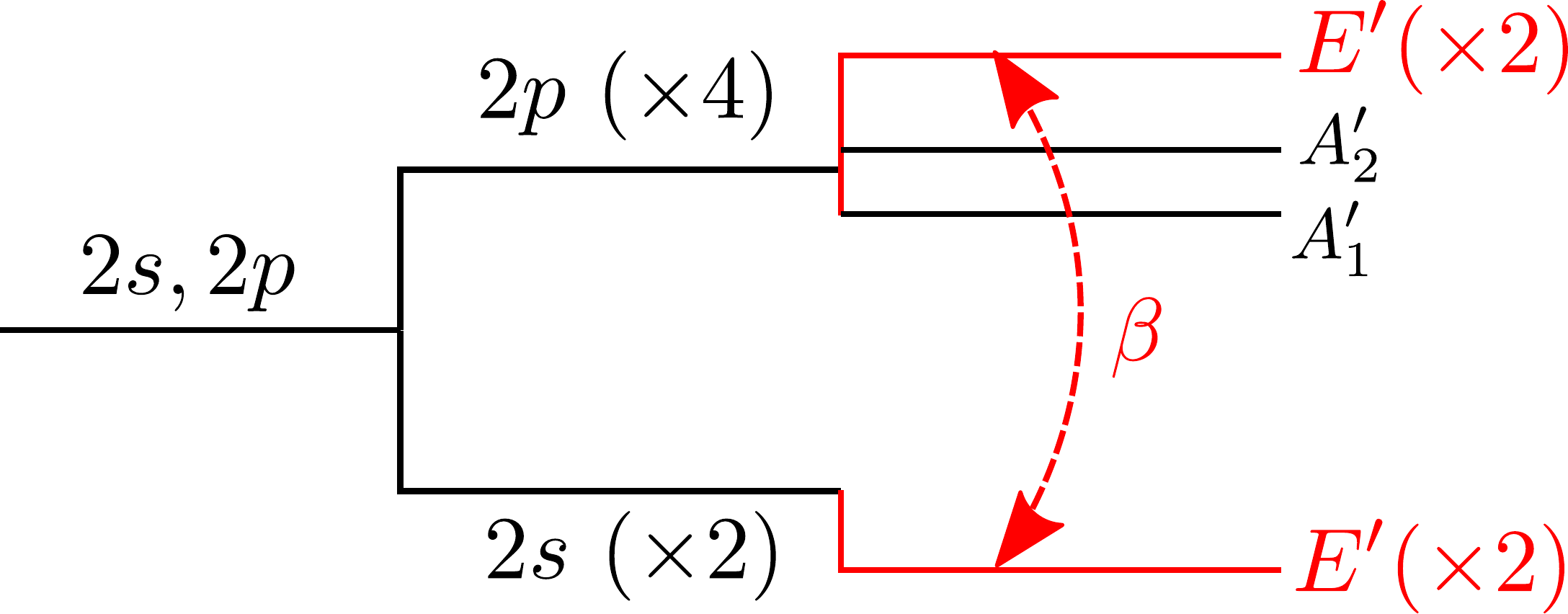}
\caption{Sketch of the $2s$ and $2p$-shell exciton fine structure. 
The splitting between the $2p$ quadruplet and $2s$ doublet (with account for the Bloch functions orbital degeneracy) arises due to the specific screening of the Coulomb interaction in TMD MLs. The splitting of the $2p$ quadruplet arises in $D_{3h}$ point symmetry and the $2p$ ($E'$) states 
are mixed with the $2s$ ($E'$) states. The order of lines is arbitrary and the splittings are shown not to scale.}\label{fig:2s2p}
\end{figure}

The symmetry of exciton basic state is, in accordance with Eq.~\eqref{ex:gen}, described by the product of irreducible representations corresponding to the relative motion wavefunction and Bloch function. Noteworthy that the symmetry of the two-dimensional envelope function $\Phi_\nu(\bm \rho)$ can be either $A'_1$ or $\Gamma_1$ in the notations of Ref.~\cite{koster63} (the identical representation, $s$-shell) or $E'$ or $\Gamma_6$ (the in-plane vector components, $p$-shells). In fact, all other 2D hydrogen-like states transform by reducible representations being combinations of those two, for instance, $d$-shell excitonic states (with the angular momentum $z$-component being $\pm 2$) transform according to $E'$, because in the $D_{3h}$ point symmetry the pair of coordinates $(x,y)$ transform exactly as the second-order combinations $(2xy,x^2-y^2)$. The Bloch functions $\mathcal U_{x,y}$ of the optically active states transform according to the two-dimensional irreducible representation $E'$. Hence, with account for the Bloch function, the $s$-shell excitons transform according to the irreducible two-dimensional representation 
\begin{equation}
\label{irrep:s}
\mathcal D_s = A'_1 \times E' = E',
\end{equation} while $p$-shell excitonic states transform according to the reducible representation 
\begin{equation}
\label{irrep:p}
\mathcal D_p = E'\times E'=A'_1 + A_2' + E'.
\end{equation}
Equation~\eqref{irrep:p} demonstrates that with account for the point symmetry of TMD ML the $p$-shell quadruplet is split into two non-degenerate sublevels ($A'_1$, $A'_2$) and the two-fold degenerate level $E'$ with the eigenfunctions (the normalization constants are omitted for brevity): 
\begin{subequations}
\label{repr:2p}
\begin{align}
A'_1 &:\quad \Phi_{p_x} \mathcal U_x + \Phi_{p_y} \mathcal U_y,\\
A_2' &:\quad \Phi_{p_x} \mathcal U_y - \Phi_{p_y} \mathcal U_x,\\
E'~(1)&: \quad \Phi_{p_x} \mathcal U_y + \Phi_{p_y} \mathcal U_x, \label{E'1}\\
E'~(2)&: \quad  \Phi_{p_x} \mathcal U_x - \Phi_{p_y} \mathcal U_y \label{E'2}.
\end{align}
\end{subequations}
The splitting between the $2p$ states transforming according to $E'$ and $(A_1'+A_2')$ was discussed in Refs.~\cite{PhysRevB.92.085413,2p:1,2p:2,Macdonald}{. It has been demonstrated on the basis of effective Hamiltonian analysis~\cite{2p:1,2p:2} and microscopic calculations~\cite{Macdonald} that the  $2p$ states with angular momentum components $\pm 1$, $\Phi_{p_{\pm 1}} \propto \Phi_{p_x} \pm \mathrm i \Phi_{p_y}$ are split within a given valley, while the time-reversal symmetry implies the equality of $\Phi_{p_{\pm 1}}$ and $\Phi_{p_{\mp 1}}$ states in $\bm K_+$ and $\bm K_-$ valleys, respectively. Equation~\eqref{repr:2p} demonstrates that the doublet $(A_1'+A_2')$ is further split in TMD MLs. This} splitting between $A_1'$ and $A_2'$ as well as the mixing between $s$-shell and $p$-shell states, see below, was not addressed.
The fine structure of $p$-shell exciton states is illustrated in Fig.~\ref{fig:2s2p} for the particular example of $2p$ excitons.
The basic functions~\eqref{repr:2p} can be represented in the alternative form in the basis of the states with a given angular momentum $z$-component $\pm 1$ as $\Phi_{p_{+1}}\mathcal U_{{-1}} \pm \Phi_{p_{-1}}\mathcal U_{{+1}}$ where the top (bottom) sign corresponding to $A'_1$ ($A'_2$) irreducible representation, and $\Phi_{p_{+1}}\mathcal U_{{+1}}$, $\Phi_{p_{-1}}\mathcal U_{{-1}}$ for two basic functions of $E'$ irreducible representation. The latter functions formally correspond to the total angular momentum component being $\pm 2$ and, since the three-fold rotation leaves third angular harmonics invariant, these functions transform exactly as the states with the angular momentum components $\mp 1$, respectively. The states $A'_1$ and $A'_2$ are, respectively, symmetrized and antisymmetrized combinations of the basic functions with the total angular momentum component being $0$. 

Moreover, it follows from Eqs.~\eqref{irrep:s} and \eqref{irrep:p} that the $s$-shell excitonic states and the combinations Eqs.~\eqref{E'1}, \eqref{E'2} of the $p$-shell  states transform according to the same irreducible representation $E'$ of the $D_{3h}$ point group. Therefore, the mixing of these states is possible. In particular, the $p$-shell state combination $ \Phi_{p_x} \mathcal U_x - \Phi_{p_y} \mathcal U_y$ mixes with $\Phi_s \mathcal U_y$ and the $\Phi_{p_x} \mathcal U_y + \Phi_{p_y} \mathcal U_x$ mixes with $\Phi_s \mathcal U_x$. In the basis of the states with given components of the total angular momentum the mixing involves the pairs:
\begin{equation}
\label{pairs}
\Phi_{p_{+1}}\mathcal U_{+1} \leftrightarrow \Phi_s\mathcal U_{-1},\quad 
\Phi_{p_{-1}}\mathcal U_{-1} \leftrightarrow \Phi_s\mathcal U_{+1}.
\end{equation} 
This mixing makes $p$-shell states optically active. {Note that this mixing is absent in the axially symmetric approximation.}
For given principal quantum numbers $n$ and $m$ of the $s$- and $p$-shell states the mixing is described by a single parameter $\beta_{nm}$. 
To illustrate this mixing in more detail we focus on the states with the principal quantum numbers $n=m=2$. We note that for purely Coulomb potential the $2s$ and $2p$ levels are degenerate. This so-called ``accidental'' degeneracy is lifted for the effective electron-hole interaction potential in TMD ML which accounts for specifics of the screening. The spin-orbit interaction as well as non-parabolicity of the single-particle energy spectrum can also contribute to the $s$-$p$ exciton splitting~\cite{PhysRevB.92.085413,2p:1,2p:2,Trushin}. To the best of our knowledge this expected splitting has not been measured so far. The effective Hamiltonian of $2s/2p$ excitons can be represented as:
\begin{equation}
\label{H:2s2p}
\mathcal H_{2X} = \begin{pmatrix}
E_{2s} & 0 & 0 & 0 & \beta & 0\\
0 & E_{2s} & 0 & 0 & 0 & \beta \\
0 & 0 & E_{2p}^{(A_1')} & 0 & 0 & 0 \\
0 & 0 & 0 & E_{2p}^{(A_2')} & 0 & 0 \\
\beta & 0 & 0 & 0& E_{2p}^{E'} & 0 \\
0 & \beta & 0 & 0 & 0& E_{2p}^{E'} 
\end{pmatrix}.
\end{equation}
Here the basic functions are chosen in the following order: $2s;E'(1)$, $2s;E'(2)$, $2p;A_1'$, $2p;A_2'$, $2p;E'(1)$~$2p;E'(2)$, the diagonal energies $E_{2s}$, $E_{2p}^{A_1'}$, etc., denote exciton energies neglecting the $s$-$p$ mixing {(but, e.g., accounting for lifting of the Coulomb degeneracy and $p_{\pm 1}$ states splitting within a given valley)}, and $\beta\equiv \beta_{22}$ is the mixing parameter. Obviously, the effective Hamiltonians for the multiplets with different principal quantum numbers have the same form with different values of parameters, the mixing is symmetry allowed between any $n$ and $m$.

\section{Microscopic origin of $s$-$p$ exciton mixing}\label{sec:micro}

In this section we develop the microscopic model of the $s$-$p$ exciton mixing. The splitting of $s$-shell and $p$-shell excitonic states for the given principal quantum number $n$ has been discussed in literature in detail, see Ref.~\cite{PhysRevB.92.085413} and references therein. The $E'$ doublet of $p$-shell states and the two states $A'_1$, $A'_2$ are split as a result of the $\bm k\cdot \bm p$-interaction and the band mixing~\cite{2p:1,2p:2}. This splitting as well as the splitting between $A'_1$ and $A'_2$ $p$-shell excitons (studied for axially-symmetric quantum dots in Ref.~\cite{glazov2007a}) is allowed in the axially symmetric model and it does not rely on specific features of the TMD symmetry and band structure. By contrast, the mixing of the $p$- and $s$-shell states requires accounting for the symmetry reduction from the axial down to the three-fold symmetry. Note that in TMD MLs the bright exciton Bloch function involves the electron state in $\bm K_\pm$ valley and the hole state in the $\bm K_\mp$ valley, see Ref.~\cite{PSSB:PSSB201552211} and the text below for details. Therefore, since the mixing involves the pairs, Eq.~\eqref{pairs}, where the electron and hole in the initial and final states should swap valleys, the coupling between $p$- and $s$-shell excitons is a combined effect of the $\bm k\cdot \bm p$-mixing of the energy bands and the electron-hole interaction. Both the short- and long-range parts of the electron-hole exchange interaction may enable simultaneous intervalley transfer of the electron and the hole~\cite{glazov2014exciton,PhysRevB.89.205303}. Below we calculate two contributions to the mixing parameter $\beta=\beta_s + \beta_l$ related with both exchange interaction mechanisms: the short range contribution, $\beta_s$, in Sec.~\ref{subsec:short} and the long-range one, $\beta_l$, in Sec.~\ref{subsec:long}. 

In what follows we present the model of the exciton mixing based on the following steps. First, we introduce the envelope functions $\Phi_{\nu}(\bm \rho)$, which can be calculated within the effective mass method as, e.g., in Ref.~\cite{Chernikov:2014a}. Second, we take into account the first-order $\bm k\cdot \bm p$ mixing between the different bands, which results in the mixing of the $s$- and $p$-shell excitonic states within a given valley{, Sec.~\ref{subsec:wave}}. Third, the short- and long-range parts of the exchange interaction are taken into account to derive the mixing between the excitons in different valleys {(Secs.~\ref{subsec:short} and \ref{subsec:long})}. {Hereafter we rely on the following relation between characteristic energies in the system:
\begin{equation}
\label{scales}
E_g \gg E_b \gg \delta_l \gg E_{exch}.
\end{equation}
Here $E_g$ is the free particle band gap, $E_b$ is the exciton binding energy, $\delta_l$ is the energy splitting of the excitonic states with the same principle quantum number and different angular momentum components (e.g., $2s/2p$ splitting) related to band non-parabolicity, spin-orbit interaction, deviation of the electron-hole interaction from the pure Coulomb $1/\rho$ law. $E_{exch}$ is the characteristic energy of the exchange electron-hole interaction in the exciton. The condition~\eqref{scales} allows one to introduce rigorously the Coulomb interaction in the first-order $\bm k\cdot \bm p$ scheme. We note, however, that in TMD MLs the exciton binding energy is just a fraction (typically $1/3\ldots 1/4$) of the band gap, therefore, in some cases, the higher-order effects due to the Coulomb interaction should be taken into account. These and other effects related to large exciton binding energy, e.g., a collapse of the ground state~\cite{collapse}, are beyond the scope of this work.}

\subsection{Exciton wavefunction}\label{subsec:wave}

We are interested in the excitons formed by the single-particle Bloch states in the vicinity of the $\bm K_\pm$ edges of the Brillouin zone. Neglecting the Coulomb-interaction induced band mixing  the two-particle Bloch function $\mathcal U_\mu(\bm \rho_e, \bm \rho_h)$ can be presented as a product of the electron and hole Bloch functions. For the excitonic state active in $\sigma^+$ polarization, $\mu=+1$, the electron Bloch state corresponds to the conduction band $c$ in the $\bm K_+$ valley and the hole state in the valence band $v$ in the $\bm K_-$ valley, because the hole state is obtained from the empty state in the valence band by the time-reversal operation~\cite{ivchenko05a,PSSB:PSSB201552211}. Within the $\bm k\cdot \bm p$ model the Coulomb interaction induces the admixture of the remote bands due to the real space localization of the envelope function and, correspondingly, its spread in the reciprocal, $\bm k$-space. We resort to the standard $\bm k\cdot \bm p$ model presented in Refs.~\cite{Kormanyos:2013dq,2053-1583-2-2-022001,2053-1583-2-3-034002} which adequately  describes the $D_{3h}$ point symmetry of the TMD ML and includes several nearest bands in agreement with DFT calculations. Disregading spin and neglecting the spin-orbit interaction we choose the basis functions for the bands $c+2, c, v, v-3$ (see Ref.~\cite{2053-1583-2-2-022001} for notations) in the $\bm K_+$ valley as follows:
\begin{equation}
\label{basic:K+}
	\mathcal U_{c+2}^{(+)}={\mathcal X- \text{i} \mathcal Y \over \sqrt{2}},~~\mathcal U_{c}^{(+)}={\mathcal X+ \text{i} \mathcal Y \over \sqrt{2}},
\end{equation}
\[
\mathcal U_{v}^{(+)}=\mathcal S,~~\mathcal U_{v-3}^{(+)}={\mathcal X- \text{i} \mathcal Y \over \sqrt{2}}.
\]
Here the subscript refers to the band, the superscript denotes the valley, and the arguments are omitted for brevity; the functions $\mathcal X$, $\mathcal Y$ transform as the coordinates $x$ and $y$, respectively, the function $\mathcal S$ is invariant.
One can see that the valence band, $v$, is invariant under the point-group transformations of TMD ML, the conduction band, $c$, transforms as the state with the angular momentum $z$-component being equal to $+1$, and the remote valence and conduction bands, $v-3$ and $c+2$, respectively, transform as the states with the angular momentum component being $-1$. The $\bm K_-$ valley states are related to the $\bm K_+$ states by the time-reversal operation, therefore the orbital Bloch functions in $\bm K_-$ valley can be obtained by the complex conjugation of the functions Eq.~\eqref{basic:K+}. The effective Hamiltonian for the $K_+$ valley in the electron representation has the following form:
\begin{equation}
\label{H:e}
\mathcal H_{+} = \left( 
\begin{array}{cccc}
	E_{c+2} & \gamma_6 k_- & \gamma_4 k_+ & 0 \\
	\gamma_6^* k_+ & E_c & \gamma_3 k_- & \gamma_5 k_+ \\
	\gamma_4^* k_- & \gamma_3^* k_+ & E_v & \gamma_2 k_- \\
	0 & \gamma_5^* k_- & \gamma_2^* k_+ & E_{v-3}
\end{array}
	\right),
\end{equation}
and in the $K_-$ valley the Hamiltonian is obtained by the substitution~\cite{2053-1583-2-2-022001}
\begin{equation}
\label{subst}
	\gamma_i \to \gamma_i^*, \qquad k_\pm \to -k_\mp.
\end{equation}
Here $\gamma_2, \ldots \gamma_6$ are the constants related with the interband matrix elements of the momentum operator, $k_\pm = k_x\pm \mathrm i k_y$ are the cyclic components of the wavevector operator $\bm k = - \mathrm i \partial /\partial \bm \rho_e$ reckoned from the Brillouin zone edge. The diagonal terms $\propto k^2$ caused by the free-electron dispersion and distant bands contributions are disregarded.   Note that if we choose all orbitals ${\cal S,X,Y}$ to be real, then all $\gamma$ parameters are purely imaginary. The Hamiltonian in the hole representation is obtained from Eq.~\eqref{H:e} by the time-reversal operation, see Ref.~\cite{birpikus_eng} for details. 

In the first order in the $\bm k\cdot \bm p$ mixing of the conduction and valence bands with the $c+2$ and $v-3$ bands we obtain the wavefunctions of the excitons $\Psi_{\bm 0;\nu;\pm 1}$ in the form
\begin{equation}
\label{exc:kp:1}
\Psi_{\bm 0;\nu;\pm 1} = \Phi_\nu(\bm \rho) \mathcal U_{\pm}(\bm \rho_e,\bm \rho_h) + \left[\hat k_\mp \Phi_\nu(\bm \rho)\right] \mathcal W_{\mp}(\bm \rho_e,\bm \rho_h) + \ldots.
\end{equation} 
Here \ldots denote omitted terms related with the $\bm k \cdot \bm p$ mixing of the $c$ and $v$ bands as well as with the distant bands beyond the $c+2$, $v-3$, these terms are not important for the following, $$\left[\hat k_\mp \Phi_\nu(\bm \rho)\right] = -\mathrm i \left(\partial/\partial x \mp \mathrm i \partial/\partial y\right)\Phi_\nu(\bm \rho),$$
the two-particle Bloch functions
\begin{equation}
\label{blochU}
\mathcal U_{\pm}(\bm \rho_e,\bm \rho_h) = \mathcal U_c^{(\pm)}(\bm \rho_e) \tilde{\mathcal U}_v^{(\mp)}(\bm \rho_h),
\end{equation}
where the tilde on top means that the Bloch function is taken in the hole representation~\cite{ivchenko05a,birpikus_eng,PSSB:PSSB201552211}, and the auxiliary combinations of Bloch functions are derived within the first-order perturbation theory as
\begin{subequations}
\label{blochW}
\begin{multline}
\label{blochWe}
\mathcal W_-= \left(\frac{\gamma_6\mathcal U_{c+2}^{(+)}(\bm \rho_e)}{E_c-E_{c+2}} + \frac{\gamma_5^* \mathcal U_{v-3}^{(+)}(\bm \rho_e)}{E_c - E_{v-3}}\right)\tilde{\mathcal U}_v^{(-)}(\bm \rho_h) \\
+\mathcal U_{c}^{(+)}(\bm \rho_e)\left( \frac{\gamma_4^*\tilde{\mathcal U}_{c+2}^{(-)}(\bm \rho_h)}{E_v-E_{c+2}} + \frac{\gamma_2 \tilde{\mathcal U}_{v-3}^{(-)}(\bm \rho_h)}{E_v - E_{v-3}}\right),
\end{multline}
\begin{multline}
\label{blochWh}
\mathcal W_+(\bm \rho_e,\bm \rho_h)= -\left(\frac{\gamma_6^*\mathcal U_{c+2}^{(-)}(\bm \rho_e)}{E_c-E_{c+2}} + \frac{\gamma_5 \mathcal U_{v-3}^{(-)}(\bm \rho_e)}{E_c - E_{v-3}}\right)\tilde{\mathcal U}_v^{(+)}(\bm \rho_h) \\
- \mathcal U_{c}^{(-)}(\bm \rho_e)\left(\frac{\gamma_4\tilde{\mathcal U}_{c+2}^{(+)}(\bm \rho_h)}{E_v-E_{c+2}} + \frac{\gamma_2^* \tilde{\mathcal U}_{v-3}^{(+)}(\bm \rho_h)}{E_v - E_{v-3}}\right).
\end{multline}
\end{subequations}
Note that the first lines of Eqs.~\eqref{blochWe}, \eqref{blochWh} result from the $\bm k\cdot \bm p$-mixing of the electron states while the second lines in Eqs.~\eqref{blochWe}, \eqref{blochWh} result from the  $\bm k\cdot \bm p$-mixing of the hole states, see Appendix~\ref{app:deriv} for the details of derivation.
The functions $\mathcal W_{\pm}$ transform, under operations from $D_{3h}$ point group, as the functions with angular momentum components $\pm 1$. Equation~\eqref{exc:kp:1} illustrates that the states with the angular momentum components $\pm 1$ and $\mp 2$ are indistinguishable in $D_{3h}$ point symmetry. This expression already describes the mixing of excitonic states with different parity, particularly, $s$- and $p$-shell states within the same valley. {Note that inclusion of other bands, beyond those included in Eq.~\eqref{basic:K+} for $\bm K_+$ valley and similar ones for $\bm K_-$, yield similar additional terms in auxiliary functions, Eq.~\eqref{blochW}. From a symmetry point of view these contributions reduce to those taken into account already. Inclusion of these disregarded bands would simply renormalize the mixing constants evaluated below, but not impact our results qualitatively.  Moreover, the mixing between the valence and conduction band produces the splitting of $p_+$ and $p_-$ states within the same valley~\cite{2p:1,2p:2}, while the mixing of $s$- and $p$-excitons occurs due to the admixture of $c+2$, $v-3$ bands.}

\begin{figure}[h]
\includegraphics[width=\linewidth]{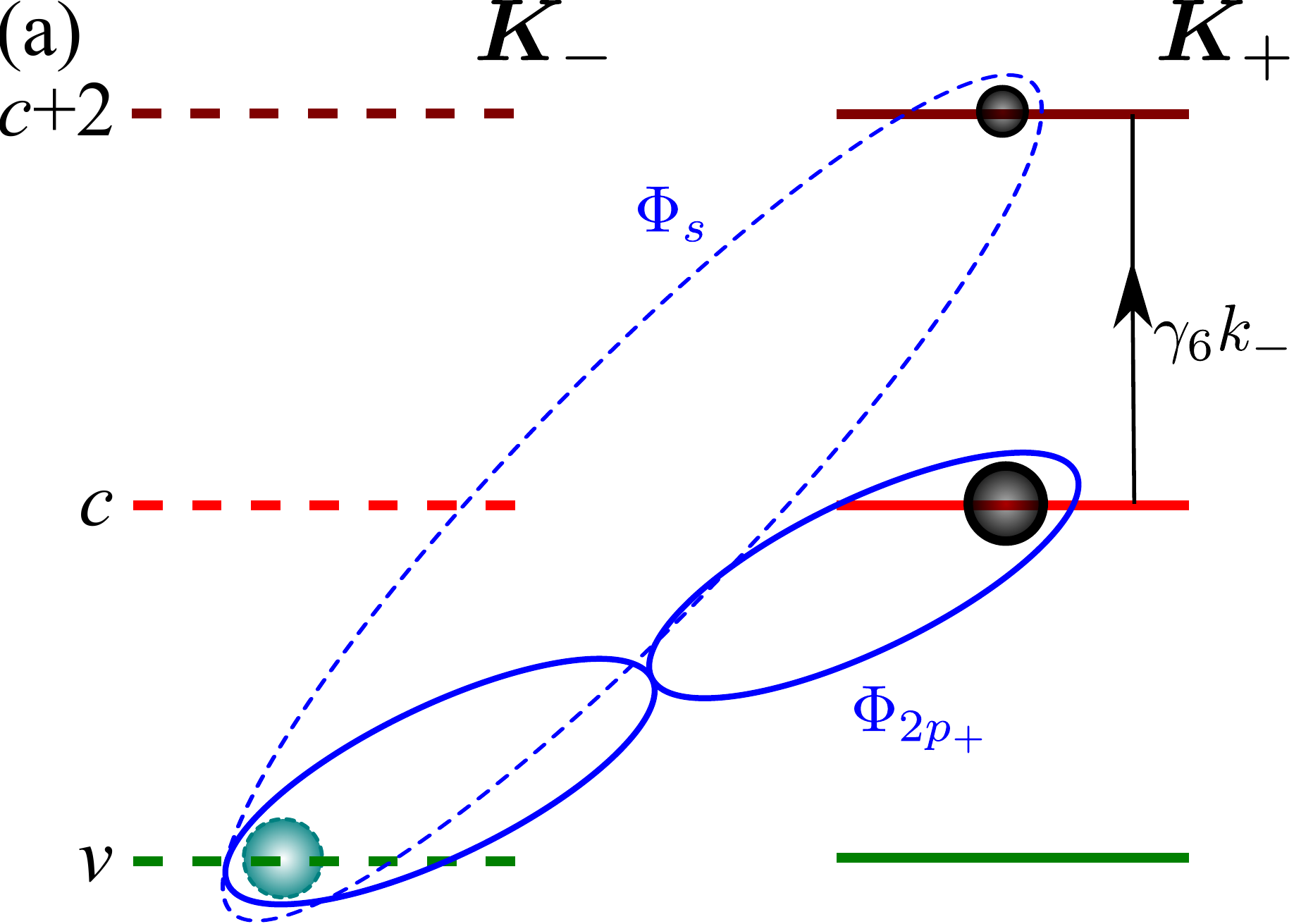}\\
\vspace{0.75cm}
\includegraphics[width=\linewidth]{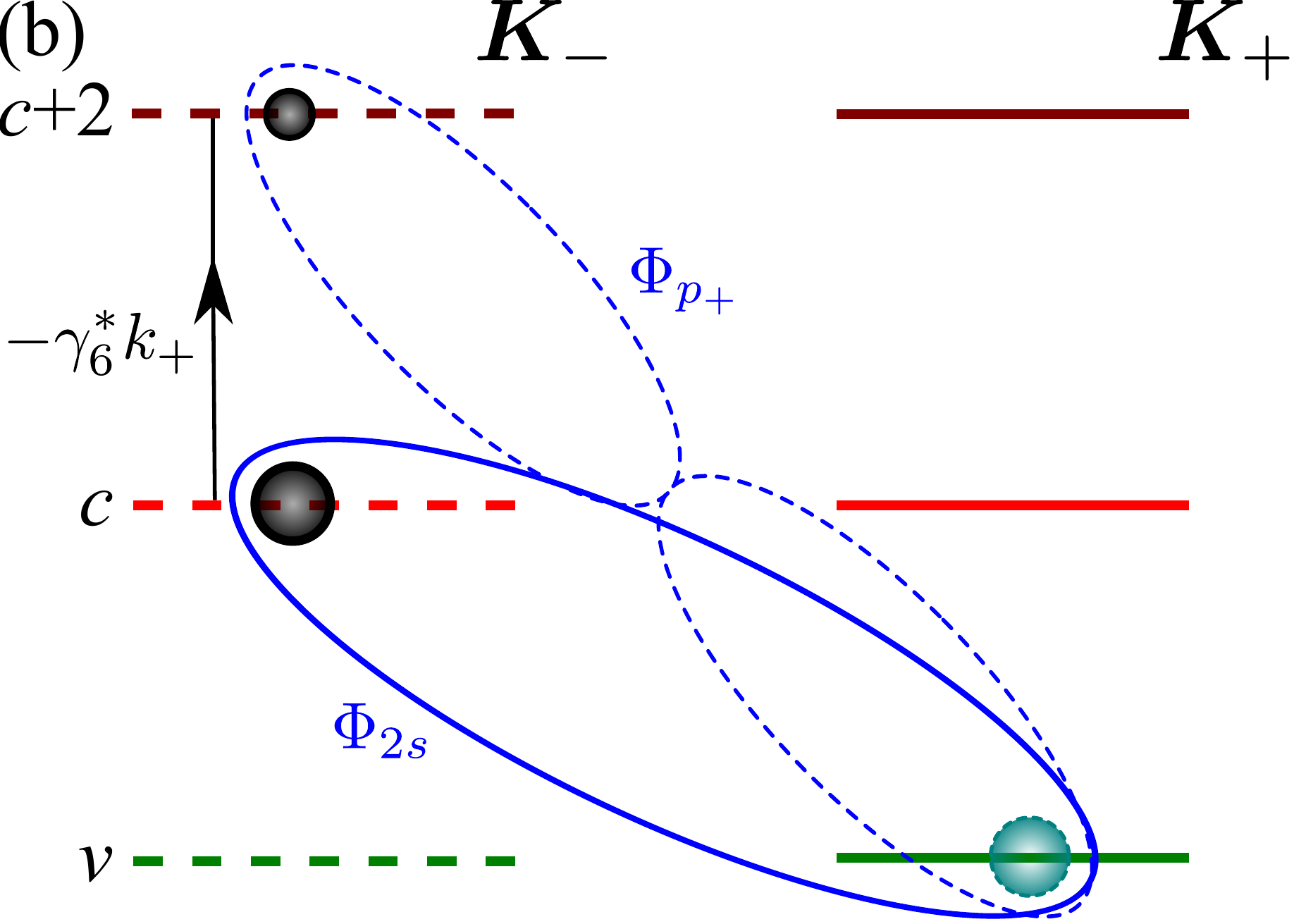}
\caption{Schematic illustration of the exciton states mixing within the same valley, Eqs.~\eqref{exc:kp:1}, \eqref{blochW}. We used here hole representation for the valence band states. (a) The admixture of $s$-shell state to $2p_+$ state in $\bm K_+$ in $\Psi_{\bm 0; 2p_+;+1}$. (b) The admixture of the $p_+$-state to the $s$-shell state in $\bm K_-$ valley, $\Psi_{\bm 0; 2s;-1}$. Black circles denote electron states, green circles denote hole states, blue solid and dashed lines denote the envelope functions: 8-like shape for $p$-shell and elliptical for $s$-shell. Vertical arrows denote the matrix elements of $\bm k\cdot \bm p$ Hamiltonians $\mathcal H_\pm$ between $c$ and $c+2$ states. }\label{fig:mixing}
\end{figure}

The mixing is schematically shown in Fig.~\ref{fig:mixing} by the example of the $2p_+$ exciton formed from the electron in the $\bm K_+$ valley, $\Psi_{\bm 0;2p_+;+1}$ (panel a), and the $2s$ exciton formed from the electron in the $\bm K_-$ valley, $\Psi_{\bm 0;2s;-1}$ (panel b). The mixing of the conduction band with the $c+2$ band is shown and the modification of the envelopes $\Phi_{2p_+} \to \Phi_s$ and $\Phi_{2s} \to \Phi_{p_+}$ described by Eq.~\eqref{exc:kp:1} is sketched. We stress that symmetry-wise the functions demonstrated on the panels (a) and (b) are equivalent.

\subsection{Short-range interaction}\label{subsec:short}

Our ultimate goal is to describe the mixing of the $\Psi_{\bm 0;2p_+,+1}$ and $\Psi_{\bm 0;2s,-1}$ where the electrons belong to different valleys. Such mixing can be enabled by the electron-hole Coulomb interaction. Particularly, its short range part is responsible for the simultaneous intervalley transfer of individual charge carriers forming an exciton. Indeed, as it follows from Eqs.~\eqref{blochW} the exciton Bloch functions combinations $\mathcal W_\pm$ transform exactly as $\mathcal U_{\pm}$ exciton Bloch functions under all $D_{3h}$ point group transformations. We present the effective Hamiltonian of the short-range electron-hole interaction in the standard form~\cite{ivchenko05a}:
\begin{equation}
\label{H:sr}
\mathcal H_{sr} = \delta(\bm \rho_e - \bm \rho_h)\, \mathcal E_0 a_0^2\, \hat V,
\end{equation}
where the Dirac $\delta$-function, $\delta(\bm \rho_e - \bm \rho_h)$, acts on the smooth envelopes, $\mathcal E_0$ and $a_0$ are the ``atomic'' energy scale and the lattice constant, respectively, and the dimensionless matrix $\hat V$ acts in the basis of two-particle Bloch functions. It has nonzero elements between the exciton Bloch functions of the same symmetry, e.g., $\langle \mathcal U_{c+2}^{(-)} \tilde{\mathcal U}_{v}^{(+)} |\hat V| \mathcal U_c^{(+)} \tilde{\mathcal U}_v^{(-)}\rangle$. Also, taking into account the spin and the spin-orbit interaction, the short-range exchange between the electron and the hole enables simultaneous flip-flop of the electron and hole spins giving rise to the mixing of $s$-shell and $p$-shell excitons of the $A$-series.

The calculation shows that the $s$-$p$ exciton mixing constant $\beta$ in the Hamiltonian~\eqref{H:2s2p} reads
\begin{equation}
\label{beta:sr}
\beta_{s}= \mathcal E_0 a_0^3   \Phi_{2s}^*(0) \Phi_{2p}'(0)  B_{s},
\end{equation}
where we introduced the notations 
\begin{equation}
\label{Phi'}
\Phi_{2p}'(\rho) = \left( {\partial \over \partial x} - \text{i}{\partial \over \partial y} \right) \Phi_{2p_+} = \left( {\partial \over \partial x} + \text{i}{\partial \over \partial y} \right) \Phi_{2p_-},
\end{equation}
and the dimensionless parameter
\begin{multline}
\label{Bsr}
B_{s} = -\frac{1 }{a_0}\left(\frac{\gamma_6\langle  \mathcal U_c^{(-)} \tilde{\mathcal U}_v^{(+)} |\hat V|\mathcal U_{c+2}^{(+)} \tilde{\mathcal U}_{v}^{(-)}\rangle}{E_c - E_{c+2}}  +\right. \\
\frac{\gamma_5^*\langle \mathcal U_{c}^{(-)} \tilde{\mathcal U}_{v}^{(+)} |\hat V| \mathcal U_{v-3}^{(+)} \tilde{\mathcal U}_v^{(-)}\rangle}{E_c - E_{v-3}} + \frac{\gamma_4^*\langle \mathcal U_{c}^{(-)} \tilde{\mathcal U}_{v}^{(+)} |\hat V| \mathcal U_c^{(+)} \tilde{\mathcal U}_{c+2}^{(-)}\rangle}{E_v - E_{c+2}}\\
\left.+\frac{\gamma_2\langle \mathcal U_{c}^{(-)} \tilde{\mathcal U}_{v}^{(+)} |\hat V| \mathcal U_c^{(+)} \tilde{\mathcal U}_{v-3}^{(-)}\rangle}{E_v - E_{v-3}}\right).
\end{multline}
Physically, the parameter $B_{s}$ accounts for the trigonality on the level of Bloch functions. 
For estimates we obtain from Eq.~\eqref{beta:sr} the short-range contribution to the mixing constant $\beta_{s} \sim \mathcal E_0 (a_0/a_B)^3B_{s}$, where $a_B$ is the exciton Bohr radius. As a result, for $\mathcal E_0=1$~eV, $a_0=3$~\AA, $a_B=15$~\AA, we have $\beta_s$ in the range from several tenths to several meV for $B_{s}$ ranging from $0.1$ to $1$. In the same way one may roughly estimate the short-range contribution to the $s$-shell exciton energy, $\sim \mathcal E_0 (a_0/a_B)^2$ is on the order of $10$~meV, in reasonable agreement with microscopic calculations~\cite{PhysRevLett.115.176801,PhysRevB.93.121107}.

\subsection{Exciton oscillator strengths and the long-range exchange interaction}\label{subsec:long}

As stated above the $s$-$p$ mixing can have a second contribution due to the long-range exchange interaction.
The $s$-shell excitons formed with electrons in $\bm K_\pm$ valley are optically active in $\sigma^\pm$ polarizations~\cite{Mak:2012qf,Sallen:2012qf}. These selection rules are described by the dipole moment operator. The reduced matrix elements of the light-matter interaction,
$ -(\hat{d}_{\pm} e_{\mp}),$ where $\hat{d}_\pm = (\hat{d}_x \pm \mathrm i \hat{d}_y)/\sqrt{2}$, $e_\pm = (e_x \pm \mathrm i e_y)/\sqrt{2}$ are the circular components of the dipole moment operator and of polarization vector, respectively, can be recast as
\begin{subequations}
\label{ns:d}
\begin{align}
&\langle \Psi_{\bm 0; ns;+ 1} |-(\hat{d}_{+} e_{-})|0\rangle = -\frac{\mathrm i\sqrt{2}e }{E_{ns}}\Phi_{ns}^*(0)\gamma_3e_{-},\\
&\langle \Psi_{\bm 0; ns;- 1} |-(\hat{d}_{-} e_{+})|0\rangle = +\frac{\mathrm i\sqrt{2} e }{E_{ns}}\Phi_{ns}^*(0)\gamma_3^*e_{+},
\end{align}
\end{subequations}
where $E_{\nu}$ is excitation energy of $\nu$th exciton. It follows from Eqs.~\eqref{exc:kp:1} and \eqref{blochW} that the $p_\pm$-shell excitons formed with electrons in $\bm K_\pm$ valley are also optically active, but in the opposite, $\sigma^\mp$, polarizations. The matrix elements of the dipole moment operator evaluated in the first-order in $\bm k \cdot \bm p$ interaction with remote $c+2$, $v-3$ bands take the form
\begin{subequations}
\label{np:d}
\begin{align}
&\langle \Psi_{\bm 0; np_+;+ 1} |-(\hat{d}_{-} e_{+})|0\rangle = -\frac{\mathrm i \sqrt{2}e}{E_{np}} \left[\Phi_{np}'(0)\right]^* \gamma_3A,\\
&\langle \Psi_{\bm 0; np_-;- 1} |-(\hat{d}_{+} e_{-})|0\rangle = +\frac{\mathrm i \sqrt{2}e}{E_{np}} \left[\Phi_{np}'(0)\right]^* \gamma_3^*A^*,
\end{align}
\end{subequations}
where
\begin{multline}
\label{a}
	A = {1\over \text{i}\gamma_3} \left( {\gamma_4\gamma_6^* \over E_{c+2}-E_c} +{\gamma_4\gamma_6^* \over E_{c+2}-E_v} \right.\\
	\left.- {\gamma_5\gamma_2^* \over E_v-E_{v-3}} -{\gamma_5\gamma_2^* \over E_c-E_{v-3}} \right).
\end{multline} 
{Note that inclusion of other distant energy bands would result in additional contributions to the parameter $A$ in Eq.~\eqref{a}.}
Equations~\eqref{np:d} can be conveniently derived taking into account the interband matrix elements of the dipole moment operator contain both $\bm k$-independent and $\bm k$-linear terms as
\begin{subequations}
\label{Mcv}
\begin{align}
&-(\bm d \cdot \bm e)^{(+)} = -\frac{\mathrm i\sqrt{2} e\gamma_3}{E_{g}} \left(e_- + \mathrm i A k_+e_+\right), \\
&-(\bm d \cdot \bm e)^{(-)} = +\frac{\mathrm i \sqrt{2}e\gamma_3^*}{E_{g}} \left(e_+ + \mathrm i A^* k_-e_-\right),
\end{align}
\end{subequations}
where the superscript $(\pm$) denotes the valley where the interband transition takes place.

Hence, for example, both $2s$-shell exciton formed with the electron in the $\bm K_+$ valley and $2p_-$-shell exciton formed with the electron in the $\bm K_-$ valley are active in the same polarization. As a result, these state become mixed via the electromagnetic field induced by the excitons. Microscopically, this mixing can be attributed to the long-range exchange interaction between the electron and hole forming the exciton~\cite{glazov2014exciton,PSSB:PSSB201552211}. Below we evaluate the mixing both in the  electrodynamical and quantum-mechanical approaches and discuss its specifics as compared with the short-range exchange contribution presented above.

{For completeness, we note that the $s$-$p$ exciton mixing due to the short-range interaction, Eq.~\eqref{beta:sr}, also produces the non-zero oscillator strength of $p$-shell excitons. The ratio of the absolute values of matrix elements for excitation of $2p$- and $2s$-excitons can be estimated as $|\beta_{s}/(E_{2p}-E_{2s})|$ for the short-range mechanism and as $|A|/a_B\sim a_0/a_B$ for the $\bm k\cdot\bm p$ interaction with remote bands. Both ratios are $\sim 0.1$ depending on the material and environment.}

In the framework of the electrodynamical approach we evaluate the interaction of the exciton with the induced electromagnetic field. As before, we focus on $2s$- and $2p$-shell states. It is instructive to consider the excitons propagating with the wavevector $\bm K\parallel x$ in the TMD ML plane excited by the plane wave with the frequency $\omega$, the wavevector components $q_x=K$, $q_y=0$,  $q_z=\sqrt{(n\omega/c)^2-K^2}$, and the field components $E_{x,y}^0$, $n$ is the refraction index of TMD ML. The exciton-induced field is found from the Maxwell equations~\cite{ivchenko05a,glazov2014exciton,PSSB:PSSB201552211}:
\begin{subequations}
\label{E:field}
\begin{equation}
\label{E:x}
E_x = E_x^0 + \frac{q_z}{q} \mathrm i G\left(\frac{|\gamma_3 \Phi_{2s}|^2}{\omega_{2s} - \omega - \mathrm i \Gamma} +  \frac{|A\gamma_3\Phi'_{2p}|^2}{\omega_{2p} - \omega - \mathrm i \Gamma}\right)E_x,
\end{equation}
\begin{equation}
\label{E:y}
E_y = E_y^0 + \frac{q}{q_z}  \mathrm i G\left(\frac{|\gamma_3\Phi_{2s}|^2}{\omega_{2s} - \omega - \mathrm i \Gamma} +  \frac{|A\gamma_3\Phi'_{2p}|^2}{\omega_{2p} - \omega - \mathrm i \Gamma}\right)E_y,
\end{equation}
\end{subequations}
where to shorten the notations the argument $\rho=0$ is omitted, $\omega_{\nu} = E_\nu/\hbar$, $\Gamma$ is the non-radiative damping of the exciton (its difference for the $s$- and $p$-shell excitons is disregarded), $q=n\omega/c$, and $G$ is the factor depending of the dielectric surrounding of the TMD ML, it can be expressed via the electromagnetic Greens function. Formally, the mechanical exciton energies in Eqs.~\eqref{E:field} should include the contributions of the short-range exchange interaction which already mixes the $s$- and $p$-shell states. The analysis of the interference of the short- and long-range exchange mechanisms is beyond the scope of the present paper. From the consistency requirement of Eqs.~\eqref{E:field} we obtain the following equations for the exciton eigenfrequencies:
\begin{multline}
\label{electrodyn:mixing}
(\omega_{2p} - \mathrm i \Gamma - \lambda_\alpha \Gamma_{0,2p})(\omega_{2s} - \mathrm i \Gamma - \lambda_\alpha \Gamma_{0,2s}) \\
= -\lambda_\alpha^2 \Gamma_{0,2s} \Gamma_{0,2p},
\end{multline}
where the subscript $\alpha=x,y$, the factors $\lambda_x= q_z/q$, $\lambda_y=q/q_z$, and we introduced the radiative decay rates of the $2s$- and $2p$-shell excitons as
\begin{equation}
\label{Gamma_0}
\Gamma_{0,2s} =  G|\gamma_3\Phi_{2s}|^2,\quad \Gamma_{0,2p} =  G|\gamma_3A\Phi_{2p}'|^2.
\end{equation}
Note that the parameter $G$ is real and positive, and $0<\Gamma_{0,2p} \ll \Gamma_{0,2s}$, the latter inequality holds because with the $\bm k\cdot \bm p$ approach the $|A|/a_B \ll 1$.

Equation~\eqref{electrodyn:mixing} demonstrates that the coupling between $s$- and $p$-shell excitons is ``dissipative'' for the states with the wavevectors within the light cone, $K< n\omega/c$. Formally this is because for real $q_z$ the right-hand side of Eq.~\eqref{electrodyn:mixing} is negative and the coupling constants $\beta_{{l,}\alpha} = \mathrm i \lambda_\alpha \hbar\sqrt{ \Gamma_{0,2s} \Gamma_{0,2p}}$ are imaginary. Physically, the excitons with $K< n\omega/c$ emit propagating electromagnetic waves, which results in the exciton damping, correspondingly, the mixing between the $2s$ and $2p$ excitons is also associated with their damping. This is quite similar to the situation of the polarization-dependent renormalization of the exciton radiative damping for the $s$-shell exciton states within the light cone~\cite{glazov2014exciton,PSSB:PSSB201552211}.

The situation is different for the excitonic states outside the light cone where $K> n\omega/c$. Such excitons do not emit propagating electromagnetic waves and $s$-$p$-shell states coupling is produced by the curl-less (longitudinal) electromagnetic field. Hence, the coupling constants $\beta_\alpha$ are real and the maximal coupling is achieved for the longitudinal exciton, $\alpha=x$:
\begin{equation}
\label{beta:x}
\beta_{l,x} = \frac{c\hbar K}{n\omega}\Gamma_{0,2s}\frac{|A\Phi'_{2p}|}{|\Phi_{2s}|}, \quad \beta_{l,y} \ll \beta_{l,x},
\end{equation}
where we assumed $K\gg n\omega/c$.
Note that if the excitons are localized as a whole by a shallow in-plane potential the mixing constant can be estimated by means of Eq.~\eqref{beta:x} replacing $K$ by the inverse localization length~\cite{glazov2007a}.
 
Equation~\eqref{beta:x} can be also derived quantum mechanically taking into account the long-range exchange interaction between the electron and the hole. To that end, let us first consider the interband scattering of two electrons from the states $c\bm k_c$ and $v\bm k_v$ to the states $v\bm k_v'$ and $c\bm k_c'$. The matrix element of this process is given by
\begin{equation}
\label{V:exch}
	V_{exch} =	 \left< \mathcal U_{c,\bm k_c'}^{(+)}(\bm \rho_2) \mathcal U_{v, \bm k_v'}^{(-)}(\bm \rho_1) \right| V_c\left|  \mathcal U_{c \bm, k_c}^{(-)}(\bm \rho_1) \mathcal U_{v, \bm k_v}^{(+)}(\bm \rho_2) \right>,
\end{equation}
where $V_c=V_c(\bm \rho_1-\bm \rho_2)$ is the Coulomb interaction between the electrons. This matrix element differs from the matrix element of the direct interaction by a substitution $\bm \rho_1 \leftrightarrow \bm \rho_2$ in the initial or final state. The specifics of TMD ML is that the interacting electrons are in different valleys. By contrast to Eqs.~\eqref{H:sr}, \eqref{Bsr} here we are interested in the long-range part of the Coulomb interaction, which allows to rearrange Eq.~\eqref{V:exch} as
\begin{multline}
\label{V:exch:1}
	V_{exch} = \left< {\cal U}_{c, \bm k_c'}^{(+)} \Big|{\cal U}_{v, \bm k_v}^{(+)}\right>
	\left< {\cal U}_{v, \bm k_v'}^{(-)} \Big|{\cal U}_{c, \bm k_c}^{(-)}\right>
	V_{c}(\bm k_c-\bm k_v')\\ \times \delta_{\bm k_c-\bm k_v',\bm k_c'-\bm k_v},
\end{multline}
where 
\begin{equation}
\label{Vc}
V_{c}(\bm K)=\frac{2\pi e^2}{n^2 K},
\end{equation}
is the Fourier image of the Coulomb interaction. Note that the screening of the long-range exchange interaction is realized by the high-frequency dielectric constant~\cite{glazov2014exciton}. The overlap integrals of the Bloch functions are non-zero if the $\bm k\cdot \bm p$ mixing of the conduction and valence bands is taken into account in the first order, cf. Eq.~\eqref{blochW}:
\begin{subequations}
\label{overlap}
\begin{align}
&\left< {\cal U}_{c, \bm k_c'}^{(+)} \Big|{\cal U}_{v, \bm k_v}^{(+)}\right>
	=   \frac{\gamma_3}{E_g} \left[K_- + \mathrm iA K_+ {(k_c'+k_v)_+ \over 2} \right], \\
&	\left< {\cal U}_{v, \bm k_v'}^{(-)} \Big|{\cal U}_{c, \bm k_c}^{(-)}\right>
	=  -\frac{\gamma_3}{E_g} \left[K_- + \mathrm iA K_+ {(k_c'+k_v)_+ \over 2} \right],
\end{align}
\end{subequations}
with $\bm K = \bm k_c'-\bm k_v = \bm k_c - \bm k_v'$.  Evaluation of Eq.~\eqref{V:exch} by virtue of Eqs.~\eqref{V:exch:1} and \eqref{overlap} and transformation to the electron-hole representation yields Eq.~\eqref{beta:x}. Note that to obtain the general expression for the $2s$-$2p$ exciton mixing for the whole range of the wavevectors including those within the light cone, one has to take into account the retardation of Coulomb interaction~\cite{goupalov98}. To that end it is sufficient to use, instead of Eq.~\eqref{Vc}, the general form of the retarded potential taken at $z=0$
\begin{multline}
V_c=\int_{-\infty}^{\infty} \frac{dk_z}{2\pi} \frac{4\pi e^2}{{n^2} [K^2+k_z^2 - (n\omega/c)^2]} \\
= \frac{2\pi e^2}{n^{2}\sqrt{K^2-(n\omega/c)^2}}.\nonumber
\end{multline}

For the states within the light cone, $K\to 0$, the efficiency of the dissipative $s$-$p$ mixing can be estimated according to Eqs.~\eqref{electrodyn:mixing} and \eqref{Gamma_0} as $|\beta_{l}|\sim \hbar \Gamma_{0s} |A|/a_B \sim \hbar \Gamma_{0s} a_0/a_B \sim 0.1\ldots 1$~meV depending on the particular material~\cite{2016arXiv161002695R} and dielectric environment which is close to the estimates in Sec.~\eqref{subsec:short} for the short-range mixing mechanism. The coupling constant is enhanced by the factor $K/(n\omega/c)>1$ for the states outside of the light cone, see Eq.~\eqref{beta:x}.

\section{Nonlinear optical properties of excitons in TMD ML$\mbox{s}$}

In this section we address the consequences of the three-fold symmetry of TMD ML, particularly, the intrinsic exciton mixing on the excitonic nonlinear optical properties, namely, in the two-photon absorption and in the second harmonic generation. It is commonly accepted that the $s$-shell exciton states are active in single-photon absorption, while $p$-shell states are active in the two-photon processes~\cite{PhysRevB.92.085413}. The lack of spatial inversion in TMD MLs enables single-photon excitation of $p$-shell excitons and two-photon excitation of $s$-shell excitons. There are two mechanisms of the selection rules modification: (i) $\bm k$-linear terms in the interband dipole matrix elements, Eqs.~\eqref{Mcv},  and (ii) the mixing of excitonic states of different parity described by the parameter $\beta$ in Eq.~\eqref{H:2s2p}. As a result, each $s$-shell and $p$-shell exciton state become simultaneously single- and two-photon active, resulting in the second harmonic generation. Below we consider both mechanisms of the nonlinear excitonic response.

\subsection{Modification of the dipole matrix elements due to \emph{k}-linear terms}\label{subsec:shg:k}

\begin{figure}[h]
\includegraphics[width=\linewidth]{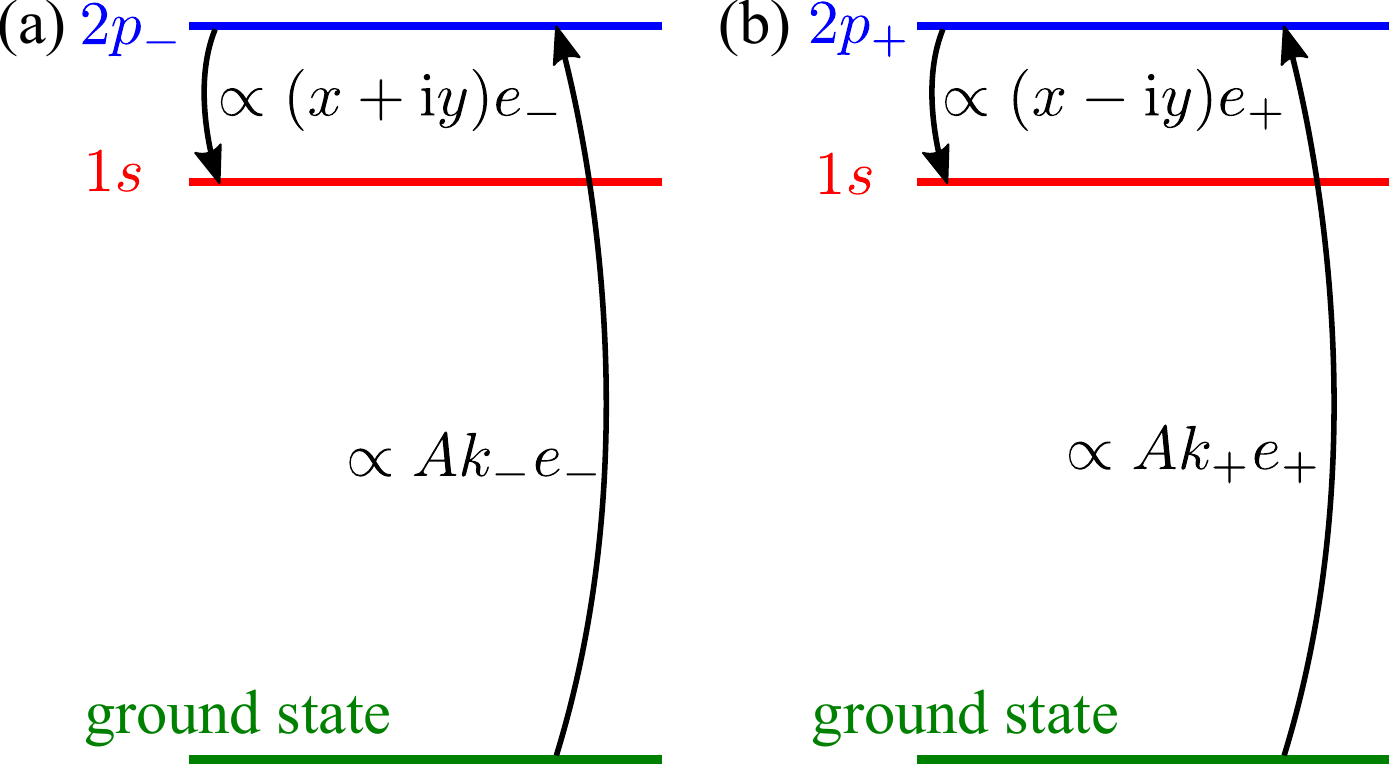}
\caption{Demonstration of the two-photon activity of the $1s$-exciton due to $\bm k$-linear terms in the interband dipole matrix element, Eq.~\eqref{Mcv}. In this mechanism the mixing of excitons is not required. Panels (a) and (b) show the exciton states with the electron in $\bm K_-$ and $\bm K_+$ valley, respectively.}\label{fig:2photonS}
\end{figure}

The physical origin of the two-photon excitation of $s$-shell excitons is illustrated in Fig.~\ref{fig:2photonS} for the particular case of the $1s$ exciton. In this two-stage process the single-photon excitation of the $2p$-exciton takes place owing to $\bm k$-linear terms in Eq.~\eqref{Mcv}. The second photon absorption goes through the inter-exciton transition~\cite{Mahan:1968ly,PhysRevB.92.085413}, which changes the envelope function parity $\langle ns| x\mp\mathrm i y| mp_\pm\rangle \ne 0$. Hence, the $s$-shell excitonic states $\Psi_{\bm 0;ns;+1}$ are excited by two $\sigma^-$ polarized photons, while the states $\Psi_{\bm 0;ns;-1}$ are excited by two $\sigma^+$ polarized photons.

Taking into account that the $\bm k$-linear terms in the interband matrix element result from the $\bm k \cdot \bm p$ interaction with remote, $c+2$ and $v-3$ bands, one can see that  the two-photon excitation of $s$-shell excitons takes place via two ``allowed'' transitions, i.e., $v\to c+2$ and $c+2 \to c$ or $v\to v-3$ and $v-3 \to c$. Under the assumptions that the exciton binding energy $E_b \ll E_g$ and that $E_v-E_{v-3}, E_{c+2} - E_c \gg E_g$ the two-photon transition matrix element can be calculated neglecting the Coulomb interaction and averaged over the exciton wavefunction afterwards. The composite matrix element of the two-photon transition takes the form~\cite{ivchenko_2phot_eng}
\begin{equation}
\label{M2phot}
M^{(2)} = \left(\frac{e}{\omega}\right)^2 \sum_m \frac{(\bm E \cdot \hat{\bm v})_{cm}(\bm E \cdot \hat{\bm v})_{mv}}{E_m - E_v - \hbar \omega},
\end{equation}
where the summation takes place over intermediate states ($m=c+2$ and $v-3$ for the $c\to v$ transition), $\hat{\bm v}$ is the velocity operator, $\bm E$ is the complex amplitude of the electric field in the incident wave. For the $c\to v$ two-photon transition at the $\bm K_+$ point one has from Eq.~\eqref{M2phot} neglecting $\hbar\omega = E_g/2$ as compared with $E_v-E_{v-3}, E_{c+2} - E_c$
\begin{equation}
\label{M2phot:a}
M^{(2,+)} = \left(\frac{e}{\omega}\right)^2 \frac{\mathrm i \gamma_3}{\hbar} A E_{\sigma^-}^2,
\end{equation}
where $E_{\sigma^\mp}$ are the corresponding circular components of the incident field and $A$ is given by Eq.~\eqref{a}. In agreement with the fact that both steps are ``allowed'' the composite matrix element, Eq.~\eqref{M2phot:a}, is independent of the wavevector. As a result, $s$-shell excitons with even envelope functions can be excited in the two-photon process. Similarly to the case of the direct interband transitions the matrix element of the two-photon excitation of the $ns$-exciton state is given by Eq.~\eqref{M2phot:a} multiplied by $\Phi_{ns}^*(0)$ (the normalization area is set to $1$):
\begin{equation}
\label{M2phot:a:exc}
M^{(2,+)}_{ns} = \left(\frac{e}{\omega}\right)^2 \frac{\mathrm i \gamma_3}{\hbar} A \Phi_{ns}^*(0) E_{\sigma^-}^2.
\end{equation}
Analogously, one can derive the expression for the matrix element $M^{(2,-)}_{ns}$ of the two-photon excitation by $\sigma^+$ polarized light, it differs from Eq.~\eqref{M2phot:a:exc} by the replacement $E_{\sigma^-}$ by $E_{\sigma^+}$ and the complex conjugation of the product $\mathrm i \gamma_3 A$.

The possibility of the two-photon excitation of $ns$ states leads to the electro-dipole second harmonic generation~\cite{pedersen}. Phenomenologically, the second harmonic generation in $D_{3h}$ point symmetry is described by the relation~\cite{valleyLPGE,Glazov2014101,PhysRevB.90.201402,Wang:2015b}
\begin{equation}
\label{phenom}
P_x^{(2\omega)} = 2\chi^{(2)} E_x E_y, \quad P_y^{(2\omega)} = \chi^{(2)} (E_x^2 - E_y^2),
\end{equation}
Here $\chi^{(2)}$ is the only allowed component of the second order susceptiblity.
For the sake of simplicity we focus on the experimentally relevant case of the second harmonic generation at $1s$ exciton. Making use of Eqs.~\eqref{M2phot:a:exc} and \eqref{ns:d} we arrive at the following expression for the susceptibility $\chi^{(2)}$ related with $\bm k$-linear terms in the dipole matrix elements:
\begin{equation}
\label{chi2:k}
	\chi^{(2)}_{k,1s} = -{e^3 |\Phi_{1s}(0)|^2\over \omega^2\omega_{1s}} A \left| {\gamma_3 \over \hbar} \right|^2 \frac{1}{2\omega - \omega_{1s} + \mathrm i \Gamma_{1s}},
\end{equation}
where $\Gamma_{1s}$ is the damping of the $1s$ exciton state.
Clearly, $\chi^{(2)}_{k,1s}$ has a resonance at $\omega= \omega_{1s}/2$ where the double frequency corresponds to the $1s$ exciton frequency in which case the real part of the denominator $2\omega - \omega_{1s} + \mathrm i \Gamma_{1s}$ vanishes.
Analogous expressions for the second-order susceptibility can be derived for $\omega$ in the vicinity of any $s$-shell excitonic resonance.

We note that the two-photon excitation of the $p$-shell excitons does not require a lack of an inversion center and takes place via a combination of the ``allowed'' interband and ``forbidden'' intraband transitions~\cite{Mahan:1968ly,PhysRevB.92.085413,note:2PAPLE}. Taking into account that the intraband matrix elements   of the velocity operator read $\bm v_{cc} = \hbar k/m_c$, $\bm v_{vv} = \hbar k/m_v$, where $m_{c,v}$ are the effective masses of the electron in the conduction and valence bands, respectively, and $1/m_c - 1/m_v = 1/\mu$ where $\mu$ is the reduced electron-hole mass we obtain from Eq.~\eqref{M2phot}
\begin{equation}
\label{M2phot:b}
\tilde M^{(2,+)} = \left(\frac{e}{\omega}\right)^2  \frac{\gamma_3}{\omega\mu}   k_+ E_{\sigma^-}^2.
\end{equation}
As compared with Eq.~\eqref{M2phot:a} for the two-step process involving intra- and interband transitions the effective two-photon matrix element is linear in the wavevector $\bm k$. That is why $p$-shell excitons are optically active in the two-photon processes and the matrix element of the two-photon excitation of the $mp$-exciton state is given, in accordance with Eq.~\eqref{M2phot:b} by
\begin{equation}
\label{M2phot:b:exc}
\tilde M^{(2,+)}_{mp} = \left(\frac{e}{\omega}\right)^2  \frac{-\mathrm i \gamma_3}{ \omega\mu} \left[\Phi_{mp}'(0)\right]^* E_{\sigma^-}^2.
\end{equation}
Due to the lack of an inversion symmetry the $p$-shell excitons are optically active in single photon processes. Taking into account Eqs.~\eqref{np:d} we have for the contribution to the susceptibility related with the $2p$-exciton 
\begin{equation}
\label{chi2:k:2p}
	\chi^{(2)}_{k,2p} = {e^3 \hbar|\Phi_{2p}'(0)|^2 \over \mu \omega^3{\omega_{2p}}} A^* \left| {\gamma_3 \over \hbar} \right|^2  \frac{1}{2\omega - \omega_{2p} + \mathrm i \Gamma_{2p}}.
\end{equation}
It is worth stressing that the susceptibility is proportional to the parameter $A$ ($A^*$) responsible for the lack of the inversion center in TMD ML.

Let us compare the contributions Eq.~\eqref{chi2:k} and \eqref{chi2:k:2p} to the nonlinear susceptibility at $1s$ and $2p$ excitons. Taking into account that $|\Phi_{2p}'(0)|^2 \sim a_B^{-2} |\Phi_{1s}(0)|^2$ and $\hbar^2/\mu a_B^2 \sim E_b$ we observe that the ratio of the peak values of $\chi^{(2)}$ can be estimated as
\begin{equation}
\label{shg:est}
\frac{\chi^{(2)}_{k,1s}(\omega=\omega_{1s})}{\chi^{(2)}_{k,2p}(\omega=\omega_{2p})} \sim \frac{E_g}{E_b} \frac{\Gamma_{2p}}{\Gamma_{1s}} \gg 1.
\end{equation}
The dominance of the $1s$ exciton contribution is related with two large factors, $\hbar\omega/E_b \sim E_g/E_b > 1$ and $\Gamma_{2p}/\Gamma_{1s}>1$ because the broadening of the excited states exceeds by far the broadening of the ground exciton state~\cite{Wang:2015b}. 

\subsection{Mixing of excitons}

\begin{figure}[h]
\includegraphics[width=\linewidth]{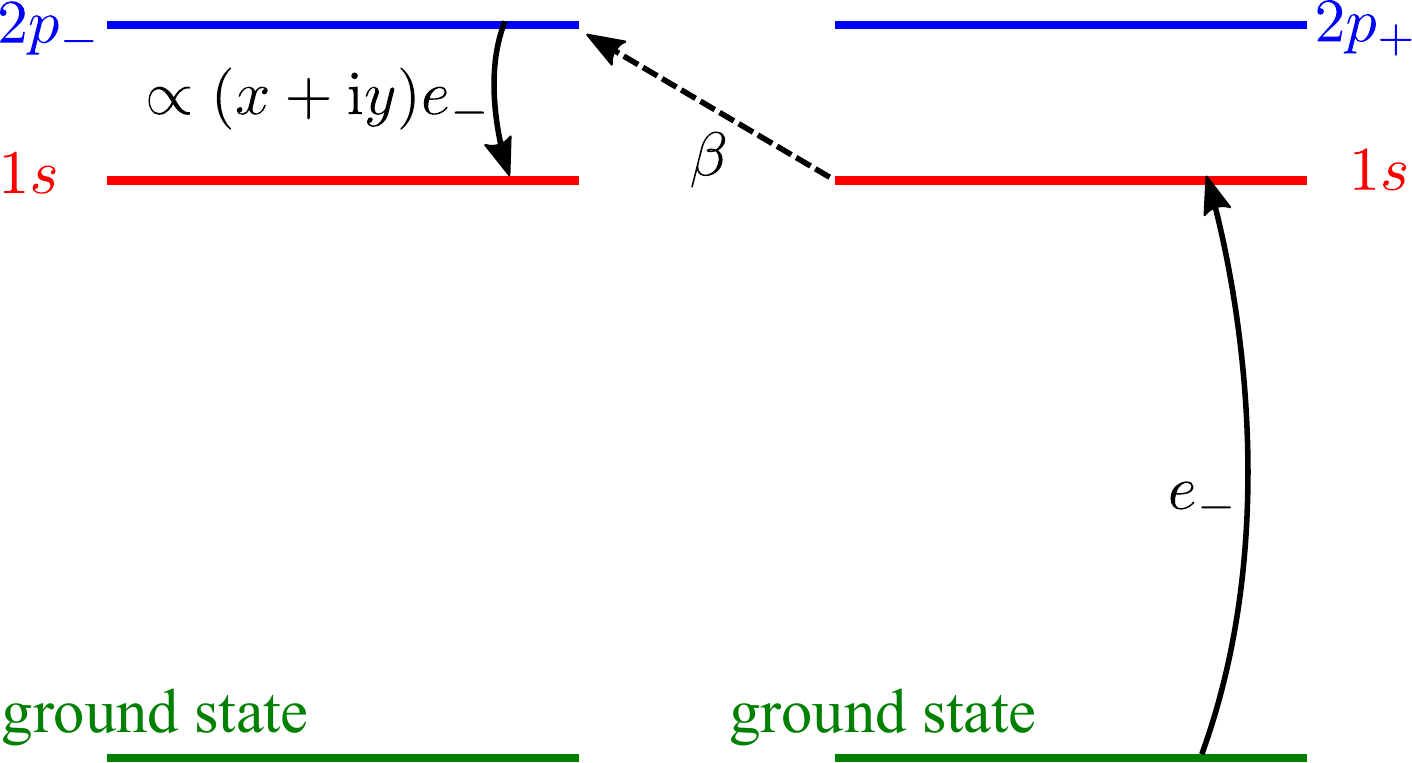}
\caption{Demonstration of the two-photon activity of the $1s$-exciton due to the $p$-$s$-exciton mixing, Eq.~\eqref{H:2s2p}. Left and right show the exciton states with the electron in $\bm K_-$ and $\bm K_+$ valley, respectively. Dashed arrows denote the mixing of excitonic states of different parity.}\label{fig:2photonBeta}
\end{figure}

The mixing of the $s$- and $p$-shell excitons described by the Hamiltonian~\eqref{H:2s2p} gives rise to additional contributions to the two-photon absorption by $s$-shell exciton states and to the second order susceptibility. These nonlinear contributions are similar to those related with the  magneto-Stark-effect or electric-field induced mixing of excitons in ZnO and GaAs~\cite{PhysRevLett.110.116402,PhysRevB.92.085202}.  Schematically, one of these contributions is illustrated in Fig.~\ref{fig:2photonBeta}: the $p$-shell exciton can be excited in a one photon process due to admixture $\propto \beta$ of the $s$-shell state, the absorption of the second photon transforms, as in the previous section, the $p$-shell exciton to the $s$-shell one.

Microscopically, the response of the TMD ML taking into account the exciton mixing can be derived similarly to the two-photon activity of the $2p$ excitons, because the two-photon excitation of the $s$-shell state takes place via a combination of the interband and intraband processes. Taking into account that the wavefunction of the $s$-shell state contains $\propto \beta$ admixture of the $p$-shell states the averaging of Eq.~\eqref{M2phot:b} produces a non-zero result for $s$-shell excitons.  Omitting the details of derivation we present the final result for the $1s$ exciton contribution to  $\chi^{(2)}$ related with the $1s$-$mp$ exciton mixing:
\begin{multline}
\label{chi2:beta}
	\chi^{(2)}_{\beta,1s} = -{e^3 \hbar \Phi_{1s}(0)\over \mu \omega^3\omega_{1s}} \left| {\gamma_3 \over \hbar} \right|^2 \sum_{m} \frac{\beta_{1m} [\Phi_{mp}'(0)]^*}{E_{1s} - E_{mp}}\times\\
	 \frac{1}{2\omega - \omega_{1s} + \mathrm i \Gamma_{1s}}.
\end{multline}
The resonance in Eq.~\eqref{chi2:beta} takes place, as in Eq.~\eqref{chi2:k}, at $\omega = \omega_{1s}/2$ when the second harmonic is resonant with the $1s$ exciton state. Similar expressions can be derived for the contributions to the nonlinear susceptibility $\chi^{(2)}$ in the vicinity of other excitonic resonances.  For instance, the second harmonic generation at $p$-shell excitons is related with the 
standard mechanism of the two-photon excitation of the $p$-shell, Eq.~\eqref{M2phot:b:exc}, and single-photon emission from the $p$-exciton due to $\propto \beta$ admixture of the $s$-shell state to the $p$-shell state.
Here lack of the inversion center is accounted for by the $s$-$p$ exciton mixing constants $\beta_{1m}$.

\subsection{Discussion and comparison with experiment}

\begin{figure}[h]
\includegraphics[width=\linewidth]{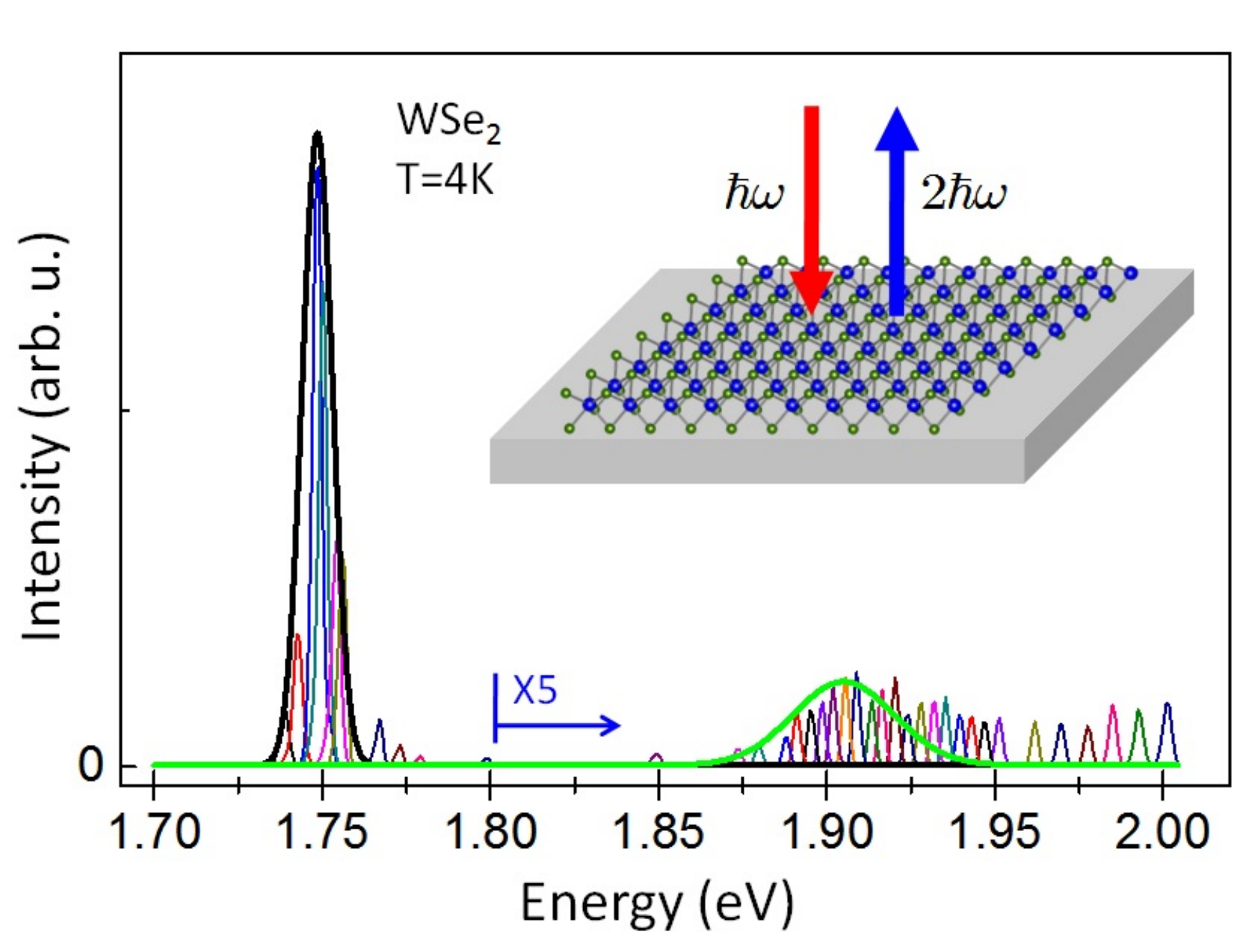}
\caption{The second harmonics generation spectrum at $T=4$~K as function of two photon energy, $2\hbar\omega$. Narrow peaks represent the emission intensity measured at a given laser frequency. Solid black and green lines represent the fits with Gauss function to the $1s$ and $2s/2p$ transition profiles. The inset shows the the schematics of the experiment. Adapted from Ref.~\cite{Wang:2015b}.}\label{fig:exper}
\end{figure}

A giant enhancement of the second harmonic generation at exciton resonances in 1 ML WSe$_2$ flakes  has been observed in Ref.~\cite{Wang:2015b}. The monolayer samples were obtained by micromechanical cleavage of bulk WSe$_2$ crystal on 90 nm SiO$_2$ on a Si substrate. For the two-photon excitation we use picosecond light pulses, generated by a tunable optical parametric oscillator synchronously pumped by a mode-locked Ti:sapphire laser.  The second harmonic emission was collected in reflection geometry, see Ref.~\cite{Wang:2015b} for details.  Figure~\ref{fig:exper} demonstrates the intensity of the second harmonic signal as a function of the doubled laser excitation energy. A strong peak at $1s$ exciton is clearly seen as well as less prominent feature at the $2s/2p$ excitons. The peaks at exciton resonances in the second-order susceptibility are in agreement with the developed model.

The dominant second harmonic response at the $1s$ exciton state can be explained within the developed model accounting for the modification of the matrix elements, Sec.~\ref{subsec:shg:k}. The estimates by Eqs.~\eqref{chi2:k},~\eqref{shg:est} show that $1s$ exciton should dominate in the second harmonic emission. As compared with all excited states the ground exciton has much smaller broadening, moreover, the $2s$ state  has much smaller oscillator strength $|\Phi_{2s}(0)/\Phi_{1s}(0)|^2 \ll 1$ already in the hydrogenic model, while as compared with $2p$ exciton the enhancement is caused by the factor $E_b/E_g$. To provide more quantitative comparison we fitted the spectra of the second harmonic generation at $1s$ and $2s/2p$ excitons by Gaussian functions (see black and green profiles in Fig.~\ref{fig:exper}) and extracted the ratio of the integral second harmonic intensities at $1s$ and $2s/2p$ states to be about 10. We abstain from more detailed fitting of the experimental data with our model due to uncertainty of the sample parameters and difficulties with evaluation of the overall emission intensity in such samples, cf.~\cite{outcoupling}. This is in reasonable agreement with the first factor in Eq.~\eqref{shg:est}, $E_g/E_b \sim 5$ in WSe$_2$.  It demonstrates that the mechanism related with the modification of selection rules, Eqs.~\eqref{chi2:k}, \eqref{chi2:k:2p}, already reproduces all main features of the experiment.

{Finally let us emphasize that the interplay of electric- and magnetic-dipole transitions may also result in the second harmonic generation due to simultaneous action of electric and magnetic fields of the light wave (so called $EB$-mechanism) as suggested in Ref.~\cite{Wang:2015b}. The $EB$-mechanism provides a nonlinear contribution with the same polarization dependence as Eq.~\eqref{phenom}, which is also enhanced at excitonic resonances. Although many parameters still need to be determined for a meaningful quantitative comparison, we expect at this stage that the $EB$-mechanism is substantially weaker in TMD MLs as compared to the ``intrinsic'' mechanism suggested here. Indeed, in order to activate the magnetic-dipole transitions at the normal incidence of radiation, the in-plane magnetic field component should play a role. However, its effect is strongly diminished in TMD MLs due to their small thickness. Here the magnetic field has to mix the states related to different bands; this effect is strongly suppressed in atomically thin systems, see, e.g., Ref.~\cite{Glazov2014101} for discussion of a similar situation realized in graphene. }

\section{Conclusion}

In the present paper we have developed the theory of the excitonic states fine structure in transition metal dichalcogenides monolayers. We have demonstrated that the lack of inversion center in the TMD MLs and their specific three-fold rotation symmetry gives rise to the mixing of the exciton states with different parity. In particular, $s$-shell and $p$-shell excitonic states whose two-dimensional envelope functions are characterized by the angular momentum components $0$ and $\pm 1$ are coupled in TMD MLs.  The microscopic theory of the mixing is developed with the $\bm k \cdot \bm p$ formalism. We have identified the origins of the mixing related with the long- and short-range parts of the electron-hole exchange interaction. The estimates show that the coupling constant for the nearest $2s$-$2p$ or $1s$-$2p$ states may be in the range of tenths to units of meV.

The manifestations of the exciton mixing in the linear and, especially, nonlinear optical properties are studied. We have demonstrated that in TMD MLs $s$-shell excitons are allowed not only in single-photon but also in two-photon processes. Similarly, $p$-shell excitons are active in single-photon absorption in addition to the two-photon absorption. We have shown that, besides exciton mixing, the lack of the inversion center in TMD MLs results in the modification of optical selection rules, which also provides simultaneous activity of $s$- and $p$-shell excitons in one- and two-photon processes. These mechanisms give rise to the effective generation of the crystallographic second harmonics in the electro-dipole approximation. The mechanism related with the modification of the selection rules provides dominant second harmonic response at $1s$-exciton resonance in qualitative agreement with experiment.

The precise evaluation of the exciton oscillator strengths and the nonlinear susceptibility by the combination of the $\bm k\cdot \bm p$ model developed here and the atomistic calculations as well as the studies of the interplay of the exciton mixing in absence and in presence of an external electric field are the tasks for future studies.

\acknowledgements

We are grateful for partial support to RFBR, the Russian Federation President Grant MD-5726.2015.2, Dynasty foundation, LIA CNRS-Ioffe RAS ILNACS, ERC Grant No. 306719, and ANR MoS2ValleyControl for financial support. X.M. also acknowledges the Institut Universitaire de France. 

\appendix

\section{Derivation of Eqs.~\eqref{exc:kp:1} and \eqref{blochW}}\label{app:deriv}

 We start from the electron-hole pair with the wavevectors $\bm k_e$ and $\bm k_h$. The Bloch function of this pair (without the Coulomb interaction) reads
\begin{equation}
\label{pair}
\Psi_{\bm k_e, \bm k_h} (\bm \rho_e, \bm \rho_h) = e^{\mathrm i \bm k_e \bm \rho_e+ \mathrm i \bm k_h \bm \rho_h} \mathcal U_{c,\bm k_e}(\bm \rho_e) \tilde{\mathcal U}_{v,\bm k_h}(\bm \rho_h).
\end{equation}
Here $\bm \rho_e$ and $\bm \rho_h$ are the in-plane position vectors of the electron and the hole, $\mathcal U_{c,\bm k_e}(\bm \rho_e)$ [$\tilde{\mathcal U}_{v,\bm k_h}(\bm \rho_h)$] are the single particle Bloch functions for the conduction band electron (valence band hole). The normalization area is set to unity. In the first order in the $\bm k\cdot \bm p$ mixing we have for the Bloch functions
\begin{align}
\mathcal U_{c,\bm k_e}(\bm \rho_e)  = \mathcal U_{c,\bm k_e=0}(\bm \rho_e) + \sum_m \mathcal U_{m,0}(\bm \rho_e) \frac{\frac{\hbar}{m_0} \langle m|\bm k_e \cdot \bm p|c\rangle}{E_c - E_m}, \label{admix:e}\\
\tilde{\mathcal U}_{v,\bm k_h}(\bm \rho_e)  = \tilde{\mathcal U}_{v,\bm k_h=0}(\bm \rho_e) + \sum_n \tilde{ \mathcal U}_{n,0}(\bm \rho_e) \frac{\frac{\hbar}{m_0} \langle n|\bm k_h \cdot \bm p|c\rangle}{E_c - E_n}. \label{admix:h}
\end{align}
The wavevectors are reckoned from the corresponding $\bm K$ point of the Brillouin zone, e.g., $\bm k_e=0$ corresponds to $\bm K_+$ valley for the electron and $\bm k_h=0$ corresponds to the $\bm K_-$ valley for the hole in the electron-hole pair excited by the $\sigma^+$ light.

Now we account for the excitonic effect. We assume that the exciton binding energy is small as compared with the bandgaps and include the Coulomb correlation in the form
\begin{multline}
\label{exciton}
\Psi_{exc} (\bm \rho_e, \bm \rho_h) =\\ \sum_{\bm k_e, \bm k_h} C_{\bm k_e, \bm k_h} e^{\mathrm i \bm k_e \bm \rho_e+ \mathrm i \bm k_h \bm \rho_h} \mathcal U_{c,\bm k_e}(\bm \rho_e) \tilde{\mathcal U}_{v,\bm k_h}(\bm \rho_h).
\end{multline}
Here $C_{\bm k_e, \bm k_h}$ is the Fourier transform of the exciton envelope function calculated within the effective mass approach,
\begin{equation}
\label{envelope}
\exp{(\mathrm i\bm K \bm R)}\Phi(\bm \rho_e-\bm \rho_h) = \sum_{\bm k_e, \bm k_h} C_{\bm k_e, \bm k_h} e^{\mathrm i \bm k_e \bm \rho_e+ \mathrm i \bm k_h \bm \rho_h}.
\end{equation}
Hence, in the first order in the $\bm k \cdot \bm p$ mixing one has
\begin{multline}
\label{exciton:1}
\Psi_{exc} (\bm \rho_e, \bm \rho_h) =\\ 
\sum_{\bm k_e, \bm k_h} C_{\bm k_e, \bm k_h} e^{\mathrm i \bm k_e \bm \rho_e+ \mathrm i \bm k_h \bm \rho_h} \mathcal U_{c,0}(\bm \rho_e) \tilde{\mathcal U}_{v,0}(\bm \rho_h) + \\
\delta \Psi_e(\bm \rho_e) \tilde{\mathcal U}_{v,0}(\bm \rho_h) + \delta \Psi_h(\bm \rho_h) \mathcal U_{c,0}(\bm \rho_e) .
\end{multline}
Making use of Eq.~\eqref{admix:e} we have for $\delta\Psi_e$:
\begin{equation}
\label{delta:e}
\delta \Psi_e(\bm \rho_e)  =
\end{equation}
\[
 \sum_{\bm k_e, \bm k_h} C_{\bm k_e, \bm k_h} e^{\mathrm i \bm k_e \bm \rho_e+ \mathrm i \bm k_h \bm \rho_h} \sum_m \mathcal U_{m,0}(\bm \rho_e) \frac{\frac{\hbar}{m_0} \langle m|\bm k_e \cdot \bm p|c\rangle}{E_c - E_m} =
 \]
\[
\sum_{\bm k_e, \bm k_h} C_{\bm k_e, \bm k_h} e^{\mathrm i \bm k_e \bm \rho_e+ \mathrm i \bm k_h \bm \rho_h} k_-^{(e)} \left[\frac{\gamma_6 \mathcal U_{c+2}^{(e)}(\bm \rho_e)}{E_c - E_{c+2}} + \frac{\gamma_5^* \mathcal U_{v-3}^{(e)}(\bm \rho_e)}{E_c - E_{v-3}} \right].
\]
In the latter equality we have considered the particular case of the electron in the $\bm K_+$ valley and took into account two admixed bands, $c+2$ and $v-3$ where
\[
\frac{\hbar}{m_0} \langle c+2|\bm k_e \cdot \bm p|c\rangle = \gamma_6 k_-^{(e)}, \quad 
\frac{\hbar}{m_0} \langle v-3|\bm k_e \cdot \bm p|c\rangle = \gamma_5^* k_-^{(e)}.
\]
To sum over $\bm k_e, \bm k_h$ in Eq.~\eqref{delta:e}  we consider the exciton with the center of mass wavevector $\bm K=0$ and make use of Eq.~\eqref{envelope} to obtain
\begin{equation}
\label{deriv}
 \sum_{\bm k_e, \bm k_h} C_{\bm k_e, \bm k_h} e^{\mathrm i \bm k_e \bm \rho_e+ \mathrm i \bm k_h \bm \rho_h} k_-^{(e)}   
\end{equation}
\[
= \sum_{\bm k_e, \bm k_h} C_{\bm k_e, \bm k_h} (-\mathrm i) \left( \frac{\partial}{\partial x_e} - \mathrm i \frac{\partial}{\partial y_e}\right) e^{\mathrm i \bm k_e \bm \rho_e+ \mathrm i \bm k_h \bm \rho_h}
  \]
  \[ 
  =  -\mathrm i \left( \frac{\partial}{\partial x_e} - \mathrm i \frac{\partial}{\partial y_e}\right)\Phi(\bm \rho_e - \bm \rho_h) =  \hat{k}_- \Phi(\bm \rho).
\]
Combining Eqs.~\eqref{exciton:1}, \eqref{delta:e} and \eqref{deriv} we obtain Eq. (10) of the manuscript with the function $\mathcal W_-$ in the form
\[
\mathcal W_-= \left(\frac{\gamma_6\mathcal U_{c+2}^{(+)}(\bm \rho_e)}{E_c-E_{c+2}} + \frac{\gamma_5^* \mathcal U_{v-3}^{(+)}(\bm \rho_e)}{E_c - E_{v-3}}\right)\tilde{\mathcal U}_v^{(-)}(\bm \rho_h).
\]
This is the first line of Eq.~\eqref{blochWe}. The second line of Eq.~\eqref{blochWe} can be derived in the same way from $\delta \Psi_h$ in Eq.~\eqref{exciton:1}.


\begin{thebibliography}{54}
\bibitem{note:2PAPLE} {For the same reason, the two-photon excitation efficiency of $s$-shell excitons is parametrically smaller than that of $p$-shell excitons. Therefore, the two-photon photoluminescence excitation is dominated by $2p$-exciton transitions. }
\end{thebibliography}
\end{document}